\input epsf
\documentclass[aps,pra,twocolumn,showpacs,superscriptaddress,letterpaper]{revtex4}

\usepackage{amsmath}
\usepackage{amssymb}
\usepackage{latexsym}
\usepackage{graphicx}

\newtheorem{theorem}{Theorem}
\newtheorem{result}{Result}

\newtheorem{proposition}[theorem]{Proposition}
\newcommand{\proof}{\noindent{\bf Proof:} }
\newcommand{\qed}{\hfill $\blacksquare$}

\newcommand{\ket}[1]{|#1 \rangle}
\newcommand{\avec}{\vec{a}}
\newcommand{\alphavec}{{\vec{\alpha}}}
\newcommand{\betavec}{{\vec{\beta}}}
\newcommand{\thetavec}{{\vec{\theta}}}
\newcommand{\gammavec}{{\vec{\gamma}}}
\newcommand{\nvec}{\vec{n}}

\newcommand{\lambdavec}{\vec{\lambda}}
\newcommand{\bs}{\beta^s}
\newcommand{\gs}{\gamma^s}
\newcommand{\bvec}{\vec{b}}
\newcommand{\sigvec}{\vec{\sigma}}
\newcommand{\XX}{X\otimes X}
\newcommand{\YY}{Y\otimes Y}
\newcommand{\ZZ}{Z\otimes Z}
\newcommand{\eqn}[1]{\label{eq:#1}}
\newcommand{\eq}[1]{Eq.~(\ref{eq:#1})}
\newcommand{\eqs}[1]{Eqs.~(\ref{eq:#1})}
\newcommand{\sgn}{\mathrm{sgn}}
\newcommand{\eig}{\mathrm{eig}}

\newcommand{\hlambda}{\varphi}
\newcommand{\hlambdavec}{\vec{\hlambda}}
\newcommand{\Leftrightarrowsp}{\hspace{0.5cm}\Leftrightarrow
\hspace{0.5cm}}
\begin{document}

\title{On the practicality of time-optimal two-qubit Hamiltonian simulation}
\author{Henry~L.~Haselgrove} %
\email{hlh@physics.uq.edu.au}
\affiliation{School of Physical
Sciences, The University of Queensland, Brisbane 4072, Australia}
\affiliation{Institute for Quantum Information, California
Institute of Technology, Pasadena CA 91125, USA}
\affiliation{Information Sciences Laboratory, Defence Science and
Technology Organisation, Edinburgh 5111 Australia}
\author{Michael~A.~Nielsen}
\email{nielsen@physics.uq.edu.au}
\homepage[\\ URL:]{http://www.qinfo.org/people/nielsen/}
\affiliation{School of Physical
Sciences, The University of Queensland, Brisbane 4072, Australia}
\affiliation{Institute for
Quantum Information, California Institute of Technology, Pasadena
CA 91125, USA}
\author{Tobias~J.~Osborne}
\email{T.J.Osborne@bristol.ac.uk}
\affiliation{School of Physical Sciences, The University of
Queensland, Brisbane 4072, Australia}
\affiliation{School of Mathematics,
University of Bristol, University Walk, Bristol BS8 1TW, United
Kingdom}
\date{\today}

\begin{abstract}
  What is the time-optimal way of using a set of control Hamiltonians
  to obtain a desired interaction?  Vidal, Hammerer and Cirac [Phys.
  Rev. Lett. \textbf{88} (2002) 237902] have obtained a set of
  powerful results characterizing the time-optimal simulation of a
  two-qubit quantum gate using a fixed interaction Hamiltonian and
  fast local control over the individual qubits. How practically
  useful are these results?  We prove that there are two-qubit
  Hamiltonians such that time-optimal simulation \emph{requires}
  infinitely many steps of evolution, each infinitesimally small, and
  thus is physically impractical. A procedure is given to determine
  which two-qubit Hamiltonians have this property, and we show that
  almost all Hamiltonians do.  Finally, we determine some bounds on
  the penalty that must be paid in the simulation time if the number
  of steps is fixed at a finite number, and show that the cost in
  simulation time is not too great.
\end{abstract}


\maketitle

\section{Introduction}

A central question of quantum information science is to determine the
minimal time required to perform a quantum computation using a set of
physical resources known to be universal for computation.  Our
understanding of what resources are universal for computation is very
well-developed, and it is
known~\footnote{See~\cite{Dodd02a,Wocjan02a,Dur01a,Bennett01a,Nielsen02d,Vidal01b,Brylinski02a,Bremner02a}
  and references therein; see also~\cite{Jones99a,Leung00a} for
  related work.} that when fast local control is available, any
unitary dynamics capable of generating entanglement is universal for
computation.  However, the question of using these resources in a
time-optimal fashion is, by comparison, understood relatively poorly.

This paper considers a particular simplified setting, that of
time-optimal simulation of two-qubit unitaries using a fixed
interaction Hamiltonian and arbitrary fast local control.
Arbitrary fast local control means that the evolution of the
interaction Hamiltonian may be interrupted by arbitrary
single-qubit operations, and that these operations take no time to
perform. This assumption corresponds to certain experimental
setups where single-qubit operations are performed on a much
faster time scale than joint operations.  Hammerer, Vidal and
Cirac \cite{Hammerer02a,Vidal02a} have given a construction for
this simulation, as well as an elegant expression for the minimum
achievable simulation time.

The simulation scheme of Hammerer {\em et al.}\ uses, in general, an
infinite number of steps to achieve time optimality. That is, the
interaction Hamiltonian is, in general, interrupted an infinite number
of times by local operations, and the time between each interruption
is infinitesimal. A simulation scheme requiring infinitely many time
steps is not practical for at least two reasons. First, the original
premise that local operations can be performed in zero time is no
longer valid if one must perform infinitely many of them.  Second, the
effects of noise on such a simulation will overwhelm the intended
coherent dynamics.  The purpose of our paper is to ask, first, whether
infinitely many time steps are actually \emph{required}, in general,
for time-optimal simulation?  We will find that the answer is yes.
Indeed, we will show that the overwhelming majority of two-qubit
interaction Hamiltonians have this property.  Given this, we then
address the question of determining how close to time optimal a
simulation can get, given that one demands a simulation using only a
finite number of time steps.

The paper is structured as follows. In
Sec.~\ref{sec:preliminaries} we review results about two-qubit
time-optimal simulation in the limit of fast control. In
Sec.~\ref{sec:procedure} we provide a procedure for determining
which two-qubit Hamiltonians require infinitesimal time steps when
used in this setting.  Finally, in Sec.~\ref{sec:finite} we
quantify the sacrifice that must be made to time-optimality when
one insists on having a simulation using a finite number of time
steps. Sec.~\ref{se:future} concludes the paper.

\section{Preliminaries} \label{sec:preliminaries}

The purpose of this section is to introduce notation and to review
some concepts and results associated with time-optimal two-qubit
simulation in the limit of fast local control. We end the section
with an introduction to the idea of a ``lazy'' two-qubit
Hamiltonian.

\subsection{Notation}

Up to rescaling of the ground state energy, an arbitrary two-qubit
Hamiltonian can be parameterised as follows:
\begin{equation}
H=I\otimes(\avec\cdot\sigvec) + (\bvec\cdot\sigvec)\otimes I +
\sum_{i,j=1}^3 M_{ij} \sigma_i \otimes \sigma_j,  \eqn{h}
\end{equation}
where $\avec\equiv(a_x,a_y,a_z)$ and $\bvec\equiv(b_x,b_y,b_z)$
are real 3-vectors, $M$ is a 3 by 3 real matrix, and
$\sigvec=(\sigma_1,\sigma_2,\sigma_3)=(X,Y,Z)$ is the vector of
Pauli operators. With respect to the computational basis $\{
|0\rangle, |1\rangle \}$, the Pauli operators are represented by
the following matrices:
\begin{equation} X=\left[\begin{array}{rr} 0&1\\1&\phantom{-}0
\end{array} \right];
\quad
 Y=\left[\begin{array}{rr} 0&-i\\i&0 \end{array}\right]
 ;\quad
 Z=\left[\begin{array}{rr} 1&0\\0&-1 \end{array}
\right].
\end{equation}
 When all the entries $M_{ij}$ are zero we say that the Hamiltonian is
{\em local}. Otherwise we say that the Hamiltonian is {\em
nonlocal}. We say that a unitary $U$ is local if it can be
expressed as a tensor product $U=A\otimes B$ of single-qubit
unitaries. Otherwise we say the unitary is nonlocal. We shall
henceforth restrict the single-qubit unitaries to be elements of
the special unitary group $SU(2)$ (i.e., the group of two-by-two
unitaries having unit determinant).

\subsection{Time optimal simulation}

A simulation scheme to approximate an arbitrary two-qubit unitary,
$U$, using a fixed Hamiltonian, $H$, and arbitrary local unitaries
may, without loss of generality, be written as follows.
\begin{eqnarray}
U&=&(A_N\otimes B_N)e^{-iHt_N}(A_{N-1}\otimes
B_{N-1})e^{-iHt_{N-1}} \nonumber \\
&& \dots(A_1\otimes B_1)e^{-iHt_1}(A_{0}\otimes B_{0}),
\label{eq:scheme}
\end{eqnarray}
where the parameters $t_n$ are nonnegative. That is, in order to
achieve the desired dynamics $U$ we can apply $H$ as many times as
we wish for arbitrary lengths of time, interspersed with arbitrary
operations on the individual qubits. We occasionally refer to
\eq{scheme} as being a {\em circuit} for $U$. It is worth noting
that the assumption that $H$ contains no $I\otimes I$ term and
that single-qubit unitaries are in $SU(2)$, implies that $U$ is in
$SU(4)$. These restrictions entail no loss in generality, as they
simply take advantage of the fact that the global phase of a
unitary operator is irrelevant.

Corresponding to the simulation \eq{scheme} is the \emph{interaction
  time}, which we define to be the total time $t_1 + \dots + t_N$ for
which the interaction Hamiltonian is applied.  For a given $U$ and
$H$, there are many possible circuits each giving rise to a
simulation of $U$. Over this range of possible circuits for $U$,
there is a corresponding range of values for the interaction time. A
circuit which achieves the minimum interaction time for a given $U$
and $H$ is said to be \emph{time-optimal}. We define $C_H(U)$ to be
the minimum achievable interaction time for simulating $U$ using $H$.
Reference \cite{Hammerer02a} gives a simple expression for $C_H(U)$,
in the two-qubit scenario. To discuss this result, we first briefly
review the canonical form of a two-qubit unitary and two-qubit
Hamiltonian operator.

\subsection{The canonical form of $U$ and $H$} \label{sec:canon}

For any unitary $U\in SU(4)$ there exists a \emph{canonical
decomposition} \cite{Khaneja01a,Kraus01a},
\begin{equation}
U=(C_1\otimes D_1)e^{-i(\theta_1 \XX + \theta_2 \YY + \theta_3 \ZZ
)}(C_2\otimes D_2), \eqn{canon}
\end{equation}
where $C_1$, $D_1$, $C_2$, and $D_2$ are single-qubit special
unitaries, and $\theta_1$, $\theta_2$, and $\theta_3$ are unique real
numbers satisfying
\begin{equation}
\pi/4 \ge \theta_1 \ge \theta_2 \ge |\theta_3| \geq 0. \label{eq:order}
\end{equation}
Although 15 parameters are needed in order to completely specify
an arbitrary two-qubit unitary $U\in SU(4)$, the canonical
decomposition shows us that the {\em nonlocal} behaviour of $U$
can be characterised in terms of only three parameters, $\theta_1$,
$\theta_2$ and $\theta_3$. We call these three parameters the
\emph{canonical-form parameters} of $U$ and the operator
$U_\thetavec \equiv e^{-i(\theta_1 \XX + \theta_2 \YY + \theta_3
\ZZ )}$ the \emph{canonical form} of $U$.

The local parts $C_1$,$D_1$,$C_2$, and $D_2$ of the canonical
decomposition do not affect the interaction time, as they can be
trivially included in the first and last steps, $A_0\otimes B_0$ and
$A_N \otimes B_N$, of a simulation. Therefore the canonical-form
parameters are all we need to know about $U$ in order to calculate the
minimum required interaction time for \eq{scheme}.

How does one calculate the canonical-form parameters? For
completeness, we review the method given in Appendix A
of~\cite{Hammerer02a}.

In the following it will be helpful to take advantage of
properties of the so-called \emph{magic basis}~\cite{Hill97a},
\setlength{\arraycolsep}{0.05cm}
\begin{eqnarray}
\ket{\mathbf{1}}&=& -\frac{i}{\sqrt{2}}(\ket{01}+\ket{10})
\nonumber \\
\ket{\mathbf{2}}&=&\phantom{-}
\frac{1}{\sqrt{2}}(\ket{00}+\ket{11})
\nonumber \\
\ket{\mathbf{3}}&=& -\frac{i}{\sqrt{2}}(\ket{00}-\ket{11})
\nonumber \\
\ket{\mathbf{4}}&=&\phantom{-}
\frac{1}{\sqrt{2}}(\ket{01}-\ket{10}). \nonumber
\end{eqnarray}
It is known~\cite{Hill97a} that, when expressed in the magic basis,
local two-qubit special unitaries are real, and canonical-form
unitaries are diagonal. This means that in the magic basis the
canonical decomposition looks like $U=RDS$, where $R$ and $S$ are real
orthogonal matrices, and $D$ is diagonal. The diagonal elements of $D$
can be easily written in terms of the canonical-form parameters of
$U$: if we define
\begin{eqnarray}
\hlambda_1&=& \phantom{-}\theta_1 + \theta_2 - \theta_3, \nonumber \\
\hlambda_2&=& \phantom{-}\theta_1 - \theta_2 + \theta_3, \nonumber \\
\hlambda_3&=&- \theta_1 + \theta_2 + \theta_3, \nonumber \\
\hlambda_4&=& -\theta_1 - \theta_2 - \theta_3,  \eqn{lambdaj}
\end{eqnarray}
then the diagonal elements of $D$ are $ e^{-i\hlambda_1}$ ,
$e^{-i\hlambda_2}$ , $e^{-i\hlambda_3}$, and $e^{-i\hlambda_4}$.
Note that \eq{lambdaj} together with \eq{order} implies that
\begin{equation}
\frac{3\pi}{4}\ge\hlambda_1 \ge \hlambda_2 \ge \hlambda_3 \ge
\hlambda_4 \ge -\frac{3\pi}{4}. \eqn{lambdaorder}
\end{equation}

We have that $U^TU=S^TDR^TRDS=S^TD^2S$, so the eigenvalues of
$U^TU$ are just the squares of the diagonal elements of $D$. That
is,
\begin{equation}
\eig(U^TU)=\{ e^{-2i\hlambda_1} , e^{-2i\hlambda_2} ,
e^{-2i\hlambda_3}, e^{-2i\hlambda_4} \}.  \eqn{eigutu}
\end{equation}

To determine the canonical-form parameters for a particular $U$, we
first calculate the eigenvalues of $U^TU$ (where the transpose is
taken in the magic basis), then derive the $\hlambda_j$ via
\eq{eigutu}, and finally solve \eqs{lambdaj}. A word of caution: the
task of deriving the $\hlambda_j$ from the $e^{-2i\hlambda_j}$ is not
as trivial as it may first seem. The problem is that, in general,
there is no guarantee that the values $-2\hlambda_j$ will lie in any
particular branch of the logarithm function. So, naively taking the
\emph{argument} of $e^{-2i\hlambda_j}$ will not necessarily give you
$-2i\hlambda_j$. A relatively simple procedure exists to correct for
this problem~\cite{Childs03a}. However in the context of
Sec.~\ref{sec:procedure} we will see later that taking the logarithm
of $e^{-2i\hlambda_j}$ along the standard branch of the logarithm
function will suffice to evaluate $-2\hlambda_j$.

Closely related to the canonical form for a two-qubit unitary is the
canonical form for a two-qubit Hamiltonian. It is discussed in section
V.A of \cite{Bennett02a}, where it is referred to as the \emph{normal}
form. Given the purely nonlocal part
\begin{equation}
H_I=\sum_{i,j=1}^3 M_{ij} \sigma_i \otimes \sigma_j
\end{equation}
of a Hamiltonian $H$, when it is expressed in the form \eq{h}, the
canonical form of $H$ is defined to be the unique Hermitian
operator
\begin{equation}
H_\alphavec=\alpha_1 \XX + \alpha_2 \YY + \alpha_3 \ZZ
\eqn{hcanon}
\end{equation}
that satisfies
\begin{equation}
H_I=(A\otimes B) H_\alphavec (A^\dagger \otimes B^\dagger)
\end{equation}
for some local unitary $A\otimes B$, where $\alpha_1 \ge \alpha_2
\ge |\alpha_3|$. The existence and uniqueness of this canonical
form is established in \cite{Bennett02a}, where it is shown that
$\alpha_1$, $\alpha_2$ and $|\alpha_3|$ are the singular values of
the matrix $M$, and $\sgn(\alpha_3)=\sgn(\det M)$.

The canonical form of a Hamiltonian $H$ encapsulates the nonlocal
behaviour of the evolution of $H$ for very small time steps. This
can be seen as follows. From sections III.B and V.A of
\cite{Bennett02a} we can write
\begin{eqnarray}
e^{-iHt}&=&(A\otimes B) e^{-iH_\alphavec t} (C \otimes D) + O(t^2)
\nonumber \\
&=& (A\otimes B) e^{-it(\alpha_1 \XX + \alpha_2 \YY + \alpha_3 \ZZ
)} (C \otimes D)\nonumber\\
&& + O(t^2)
\end{eqnarray}
for some local unitaries $A$,$B$, $C$, and $D$. To order $t$, the
evolution of $H$ is given by a unitary having canonical-form
parameters $t\alpha_1$, $t\alpha_2$ and $t\alpha_3$.

\subsection{Expression for $C_H(U)$} \label{sec:chu}
We are almost ready to review the expression for $C_H(U)$ given in
\cite{Vidal02a,Hammerer02a}. Before we do so we review the concept of
\emph{special-majorization}. Special-majorization describes a
particular type of partial ordering of three-vectors. Its use in
\cite{Vidal02a,Hammerer02a} allows certain results to be described very
succinctly.  To define special-majorization, it is necessary to first
introduce the idea of a \emph{special-ordered} three-vector.  Given a
real vector $\betavec = (\beta_1,\beta_2,\beta_3)$, the corresponding
\emph{special-ordered} vector $\betavec^s$ is defined as follows. The
absolute value of the components of $\betavec^s$ are given by the
absolute value of the components of $\betavec$ rearranged in
nonincreasing order. That is, $|\beta^s_j|=|\beta_{\pi(j)}|$ for the
permutation $\pi(j)$ that gives $|\beta^s_1| \ge |\beta^s_2| \ge
|\beta^s_3|$. The definition is completed by specifying that
$\beta^s_1$ and $\beta^s_2$ are nonnegative, and that $\beta^s_3$ has
the same sign as the product $\beta_1\beta_2\beta_3$. Then, $\betavec$
is said to be special-majorized by $\gammavec$ (denoted $\betavec
\prec_s \gammavec$) if
\begin{eqnarray}
\bs_1 &\le& \gs_1 \nonumber\\
\bs_1+\bs_2-\bs_3 &\le& \gs_1 + \gs_2 - \gs_3 \\
\bs_1+\bs_2+\bs_3 &\le& \gs_1 + \gs_2 + \gs_3, \nonumber \\
\end{eqnarray}
where $\betavec^s$ and $\gammavec^s$ are the special-ordered
versions of $\betavec$ and $\gammavec$.

Let $H$ be a two-qubit Hamiltonian having canonical form
$H_\alphavec$ and let $U$ be a two-qubit unitary having canonical
form $U_\thetavec$. Then $C_H(U)$, the minimum time required to
simulate $U$ using $H$, is given by the minimum value of $t_S \ge
0$ such that either
\begin{eqnarray}
\thetavec &\prec_s& \alphavec t_S \hspace{3em}\mathrm{or} \\
 \thetavec
+ (-\frac{\pi}{2},0,0) &\prec_s& \alphavec t_S
\end{eqnarray}
holds.

\subsection{The lazy Hamiltonian} \label{sec:lazy}

We now introduce the the central concept of a ``lazy'' Hamiltonian.
For a given two-qubit Hamiltonian $H$, we define a function $\tau(t)$
as follows:
\begin{equation}
\tau(t)=C_H(e^{-iHt}).
\end{equation}
That is, $\tau(t)$ is the minimum total time for which the Hamiltonian
$H$ must be applied, when it is being used together with arbitrary
local unitaries, to simulate its own action $e^{-iHt}$.
Such a simulation would be of the form
\begin{eqnarray}
e^{-iHt}&=&(A_N\otimes B_N)e^{-iHt_N}(A_{N-1}\otimes
B_{N-1})e^{-iHt_{N-1}} \nonumber \\
&& \dots(A_1\otimes B_1)e^{-iHt_1}(A_{0}\otimes B_{0}).
\eqn{selfsim}
\end{eqnarray}

The trivial ``simulation'' having the single step $e^{-iHt}$ has
an interaction time of $t$. Thus the minimum achievable
interaction time will be no greater than $t$:
\begin{equation}
\tau(t)\le t.
\end{equation}
Under what circumstances will $\tau(t)$ be less than $t$? It turns
out that this question is very closely linked to our main
question: what are the circumstances under which a time-optimal
simulation will require infinitesimal time steps?

Consider the class of two-qubit Hamiltonians having the following
property:
\begin{equation}
\tau(t)<t \hspace{4em} \forall t>0.   \label{eq:lazy}
\end{equation}
We shall say that a Hamiltonian is ``lazy'' if it is nonlocal and
satisfies \eq{lazy}.
\begin{proposition} {} \label{prop:lazy}
If a Hamiltonian is lazy, then the time-optimal simulation of any
two-qubit nonlocal unitary $U$ using $H$ requires infinitely many
time steps.
\end{proposition}
\proof Suppose there exists a time-optimal simulation scheme, of the
form of \eq{scheme}, for a nonlocal $U$ using a lazy Hamiltonian $H$,
where the number of time steps $N$ is finite. At least one of the
$t_n$ must be nonzero, otherwise the simulation would be unable to
produce nonlocal dynamics. For such a nonzero $t_n$, consider the
corresponding factor $e^{-iHt_n}$ in the simulation. Since $H$ is
lazy, there exists a simulation for $e^{-iHt_n}$ having an interaction
time {\em less} than $t_n$. If we substitute such a simulation for
$e^{-iHt_n}$ back into the simulation for $U$, then the new simulation
for $U$ now has a lesser interaction time than it did before. However
this contradicts the assumption that the original simulation was time
optimal. Hence the premise that the original simulation had finite
time steps is false. Thus we conclude that any lazy Hamiltonian will
require infinitesimal time steps when used for the time-optimal
simulation of any nonlocal two-qubit unitary $U$.  \qed

To show that a particular $H$ is lazy, it is sufficient to show that
$\tau(t)<t$ for all $t$ in some interval $(0,\epsilon)$ for any
positive $\epsilon$. To see this, note that if there is a simulation
for $e^{-iHt}$ with interaction time $t_s<t$, then clearly there
exists a simulation for $e^{-iHnt}$ with interaction time $t_sn$, for
any positive integer $n$. Thus, $\tau(t)<t$ implies that $\tau(nt)<nt$
for all positive integers $n$. So, if $\tau(t)<t$ for all
$t\in(0,\epsilon)$, then $\tau(t)<t$ for all $t>0$.

\section{General Procedure} \label{sec:procedure}

Which two-qubit Hamiltonians are lazy?  We have seen in the previous
section (proposition~\ref{prop:lazy}) that lazy two-qubit Hamiltonians
require infinitely many time steps if they are to be used for
time-optimal control, and thus are impractical.  In this section we
provide a simple set of sufficient conditions for a Hamiltonian to be
lazy, expressed in terms of the parameters of the Hamiltonian. The
parameterisation in \eq{h} is more general than it needs to be for
this purpose. We can simplify matters by using the fact that a
Hamiltonian $H$ is lazy if and only if $( A \otimes B) H (A^\dagger
\otimes B^\dagger)$ is lazy, where $A$ and $B$ are any single-qubit
unitaries. This is a consequence of the fact that $e^{-iHt}$ has the
same canonical form as $e^{-i ( A \otimes B) H (A^\dagger \otimes
  B^\dagger) t}$.  Thus, without loss of generality we choose to only
consider Hamiltonians where the purely nonlocal part is in canonical
form, that is
\begin{equation}
H=I\otimes(\avec\cdot\sigvec) + (\bvec\cdot\sigvec)\otimes I +
\sum_{j=1}^3 \alpha_j \sigma_j \otimes \sigma_j,  \eqn{hs}
\end{equation}
where $\alpha_1\ge \alpha_2\ge|\alpha_3|$.

Recall from subsection \ref{sec:lazy} that we define a Hamiltonian to
be lazy if $\tau(t)<t$ over some interval $(0,\epsilon)$.  Suppose we
could find a Taylor series expansion
\begin{equation}
\tau(t)=\tau^{(0)} + \tau^{(1)}t + \tau^{(2)}t ^2 + \dots
\end{equation}
for $\tau(t)$ in the variable $t$. Thus,
\begin{equation}
\tau(t)-t=\tau^{(0)} + (\tau^{(1)}-1)t + \tau^{(2)}t ^2 + \dots
\eqn{taut}
\end{equation}
Then, because we can assume $t$ is small, the corresponding
Hamiltonian is lazy if and only if the first nonzero item in the list
$\tau^{(0)}$, $(\tau^{(1)}-1)$, $\tau^{(2)}$, $\tau^{(3)}$, \dots {}
is negative. Our procedure involves finding expressions for the first
few items in that list, in terms of the parameters $\avec$, $\bvec$
and $\alphavec$ of the Hamiltonian. We then find the conditions under
which each expression will be negative. We find that in fact
$\tau^{(0)}=\tau^{(1)}-1=\tau^{(2)}=0$ always, and so $\tau^{(3)}$ is
the first term that may be negative.  Accordingly, in the analysis
that follows we consider the behaviour of $\tau(t)$ up to order $t^3$,
so as to arrive at some nontrivial conditions for a Hamiltonian being
lazy.

\subsection{Procedure to find the Taylor coefficients of $\tau(t)$}

We seek expressions for the Taylor coefficients of $\tau(t)$, namely,
$\tau^{(0)}$, $\tau^{(1)}$, $\tau^{(2)}$, and $\tau^{(3)}$. The
expression for $C_H(U)$ involves the canonical-form parameters
$\theta_1$, $\theta_2$ and $\theta_3$ of the unitary $U$. So we first
try to find expressions for the canonical-form parameters
$\theta_1(t)$, $\theta_2(t)$ and $\theta_3(t)$ of the unitary
$e^{-iHt}$. From subsection \ref{sec:canon}, the canonical-form
parameters can be expressed in terms of parameters $\hlambda_1$,
\dots, $\hlambda_4$, where
\begin{equation}
\eig( e^{-iH^Tt}e^{-iHt} )=\{
e^{-i2\hlambda_1},\dots,e^{-i2\hlambda_4} \},
\end{equation}
with the transpose taken in the magic basis. Thus,
\begin{equation}
\hlambdavec(t)=\left( -\frac{1}{2} \arg(\eig( e^{-iH^Tt}e^{-iHt}
)) + \nvec \pi \right)^\downarrow
\end{equation}
where the vector of integers $\nvec$ accounts for the ambiguity in
taking the argument, and where the down-arrow sorts in decreasing
order so that we are in agreement with the ordering of the
$\hlambda_j$ in \eq{lambdaorder}. However, since we are only
interested in the behaviour over a small interval $t\in[0,\epsilon]$,
it turns out we can take $\nvec=0$. This can be seen as follows. From
the discussion at the end of subsection \ref{sec:chu}, for small $t$
the canonical-form parameters of $e^{-iHt}$ will be small. Thus, the
parameters $\hlambda_1,$\dots$, \hlambda_4$ will also be small. But
this can only be the case when $\nvec=0$.  (A more rigorous proof of
this fact is easily deduced from the procedure for finding the
canonical-form parameters described in~\cite{Childs03a}.) Thus,
 \begin{equation}
\hlambdavec(t)=\left( -\frac{1}{2} \arg(\eig( e^{-iH^Tt}e^{-iHt}
)) \right)^\downarrow
\end{equation}
for $t$ in some interval $[0,\epsilon]$.

Now, it is possible to write
\begin{equation}
e^{-iH^Tt}e^{-iHt}=e^{K(t)},
\end{equation}
where $K(t)$ is given by the Campbell-Baker-Hausdorf series (for a
derivation see, for example, \cite{Wilcox67a})
\begin{eqnarray}
K(t)&=&-it(H^T+H)+\frac{1}{2}(-it^2)[H^T,H] + \nonumber \\
&&\frac{1}{12}(-it)^3([H^T,[H^T,H]]+[H,[H,H^T]])\nonumber\\
&& + \dots, \eqn{CBH}
\end{eqnarray}
where $[A,B]=AB-BA$. Thus, for $t \in [0,\epsilon]$ we can write
\begin{eqnarray}
\hlambdavec(t)&=&\left( -\frac{1}{2} \arg(\eig(e^{K(t)} ))
\right)^\downarrow \\
&=& \left(-\frac{1}{2i} \eig(K(t) ) \right)^\downarrow.
\end{eqnarray}

{}From Theorem~1.10 of \cite{Kato82a}, a normal-valued operator
function that can be expressed as a power series
\begin{equation}
T(t)=T^{(0)}+T^{(1)}t+T^{(2)}t^2 + \dots
\end{equation}
has eigenvalues which are holomorphic functions of $t$. Thus, the
entries of the vector $-\frac{1}{2i} \eig(K(t) )$ can be expressed as
holomorphic functions of $t$, and the components of $\hlambdavec(t)$
are therefore continuous piecewise-holomorphic functions of $t$, over
some interval $[0,\epsilon]$. Approximating $K(t)$ to some order in
$t$ will give the same order of approximation for $\hlambdavec(t)$:
\begin{equation}
\hlambdavec(t)+O(t^n) = \left(  -\frac{1}{2i} \eig(K(t)+O(t^n))
\right)^\downarrow.
\end{equation}

Define $\tilde{K}(t)$ to be the first three terms in the expansion
of $K(t)$ in \eq{CBH}. That is,
\begin{eqnarray}
\tilde{K}(t)&=&-it(H^T+H)+\frac{1}{2}(-it^2)[H^T,H] + \nonumber \\
&&\frac{1}{12}(-it)^3([H^T,[H^T,H]]+[H,[H,H^T]]).
\end{eqnarray}

Then, if we define
\begin{equation}
\lambdavec(t)=-\frac{1}{2i} \eig (\tilde{K}(t))
\end{equation}
we have
\begin{equation}
\hlambdavec(t)=\left( \lambdavec(t)\right)^\downarrow + O(t^4).
\eqn{lambdavec2}
\end{equation}
We can find the first four Taylor coefficients of each component
of $\lambdavec(t)$ in the following way. Each component of
$\lambdavec(t)$ satisfies the characteristic equation
\begin{equation}
f(t)=\det (\tilde{K}(t) + 2i\lambda(t)) = 0.
\end{equation}
We can substitute a Taylor series for $\lambda(t)$:
\begin{equation}
f(t)=\det (\tilde{K}(t) + 2i (\lambda^{(0)} + \lambda^{(1)}t +
\dots ) ) = 0.  \eqn{dettaylor}
\end{equation}

 Since \eq{dettaylor} must be true for a range of values of $t$,
 then all coefficients in a Taylor series for $f(t)$ must be
 zero. Finding expressions for the coefficients $f^{(j)}$,
 and solving the equations $f^{(j)}=0$, will give us the coefficients $\lambda^{(0)}$,
 $\lambda^{(1)}$,\dots etc. Note that when we wish to find a
 particular term $f^{(j)}$, we need only include terms in the
 expansion of $\lambda(t)$ in \eq{dettaylor} to order $t^j$.

How can we use the Taylor coefficients of $\lambdavec(t)$ to find
the Taylor coefficients of $\hlambdavec(t)$? Does the ordering
operation in \eq{lambdavec2} present a difficulty? Not really. Say
that we knew the Taylor series for each of the four functions
$\lambda_1(t)$,\dots,$\lambda_4(t)$. We would simply order these
vector components with respect to the zero-order Taylor
coefficients: $\lambda^{(0)}_1 \ge \lambda^{(0)}_2 \ge
\lambda^{(0)}_3 \ge \lambda^{(0)}_4$. If it happened that none of
the $\lambda^{(0)}_j$ were equal, that is $\lambda^{(0)}_1 >
\lambda^{(0)}_2 > \lambda^{(0)}_3 > \lambda^{(0)}_4$, then
\eq{lambdavec2} would immediately imply that
\begin{equation}
\hlambdavec(t)=\lambdavec(t)+O(t^4) \eqn{lambdavec3}
\end{equation}
for $t$ in some interval $[0,\epsilon]$.  That is, for small $t$,
$\hlambdavec(t)$ is equal to $\lambdavec(t)$ up to order $t^3$, where
the components of $\lambdavec(t)$ are arranged so the zeroth Taylor
coefficients are in decreasing order.  In the special case where some
of the zero-order coefficients $\lambda^{(0)}_1, \lambda^{(0)}_2,
\lambda^{(0)}_3, \lambda^{(0)}_4$ are equal, then we break the tie by
considering the first-order coefficients, and if those are equal we
consider the next highest order and so-on.  In what follows, we will
use the ordering scheme as described in this paragraph, so that we may
use \eq{lambdavec3} instead of \eq{lambdavec2}.

\subsection{Procedure to find the Taylor coefficients of the
components of $\lambdavec(t)$.} %

In this subsection we describe the calculation of the Taylor
coefficients of the components of $\lambdavec(t)$.

In the magic basis, the Hamiltonian, $H$, in \eq{hs} reads
\begin{eqnarray}
&&H= \nonumber \\
&&
 \left(
\begin{array}{lccr}
\alpha_1{-}\alpha_2{+}\alpha_3 & {-}ia_3{-}ib_3 & {-}ia_2{+}ib_2 &
{-}ia_1{-}ib_1
\\
ia_3{+}ib_3 & {-}\alpha_1{+}\alpha_2{+}\alpha_3 & ia_1{-}ib_1 &
{-}ia_2{-}ib_2
\\
ia_2{-}ib_2 & {-}ia_1{+}ib_1 & {-}\alpha_1{-}\alpha_2{-}\alpha_3 &
ia_3{-}ib_3
\\
ia_1{+}ib_1 & ia_2{+}ib_2 & {-}ia_3{+}ib_3 &
\alpha_1{+}\alpha_2{-}\alpha_3
\end{array}
\right). \nonumber \\
&&
\end{eqnarray}
The components of $\tilde{K}(t)$ in the magic basis are found with the
aid of the computer algebra system Maple.  This is straightforward in
any of the standard computer algebra systems, and the specific form is
both complex and not particularly illuminating, so we will not
reproduce the components of $\tilde{K}(t)$ here.  Note that each
component is a third-order polynomial in $t$.

Next, we find expressions for $f^{(0)},\dots, f^{(6)}$. To find
$f^{(n)}$, we evaluate the expression $f(t)$ in \eq{dettaylor}
using Maple, including terms in the series $\lambda^{(0)} +
\lambda{(1)}t + \dots $ up to at least order $t^n$. Then,
$f^{(n)}$ is given by the coefficient of the $t^n$ term in this
expression. Explicit expressions for $f^{(n)}$ will not be given
here as they are rather lengthy and not illuminating. They are
polynomials in the parameters of the Hamiltonian.

The next step is to solve the equations $f^{(n)}=0$,
$n=0,\ldots,6$, via Maple. The results are as follows.
\begin{itemize}
\item Solving $f^{(0)}=0$ yields $\lambda^{(0)}=0$. That is, the
zero-order Taylor coefficients of each component of
$\lambdavec(t)$ are zero. Thus, from \eq{lambdavec3} and
\eq{lambdaj} we have $\thetavec^{(0)}=0$.

\item Solving $f^{(1)}=0$, $f^{(2)}=0$, and $f^{(3)}=0$ provides
no new information about the $\lambda^{(n)}$.

\item Solving $f^{(4)}=0$ yields four solutions, one for each
component in $\lambdavec(t)$. We write them in nonincreasing order
as follows:
\begin{eqnarray}
\lambda^{(1)}_1&=& \alpha_1+\alpha_2-\alpha_3\nonumber \\
\lambda^{(1)}_2&=& \alpha_1-\alpha_2+\alpha_3\nonumber \\
\lambda^{(1)}_3&=& -\alpha_1+\alpha_2+\alpha_3\nonumber \\
\lambda^{(1)}_4&=& -\alpha_1-\alpha_2-\alpha_3
\end{eqnarray}
This gives $\thetavec^{(1)}=(\alpha_1,\alpha_2,\alpha_3)$.

\item Solving $f^{(5)}=0$ gives $\lambda^{(2)}=0$. Thus,
$\thetavec^{(2)}=0$.

\item Solving $f^{(6)}=0$ gives four solutions to $\lambda^{(3)}$,
so long as we assume $\alpha_1 > \alpha_2 > \alpha_3$. Each of the
four solutions for $\lambda^{(3)}$ correspond to one of the four
solutions to $\lambda^{(1)}$ (which were substituted in turn).
Thus, we are able to correctly associate each of the four
solutions to a particular component of the ordered vector
$\lambdavec(t)$. For the sake of brevity we will not reproduce the
expressions for $\lambdavec^{(3)}_j$. Rather, we just provide the
resulting expressions for $\theta^{(3)}_j$:
\begin{eqnarray}
\theta^{(3)}_1&=&\frac{1}{6}\left( -\alpha_1
(a_2^2+a_3^2+b_2^2+b_3^2)+2\alpha_2a_3b_3 + 2\alpha_3a_2b_2
\right) \nonumber\\
\theta^{(3)}_2&=&\frac{1}{6}\left( -\alpha_2
(a_1^2+a_3^2+b_1^2+b_3^2)+2\alpha_1a_3b_3 + 2\alpha_3a_1b_1
\right) \nonumber \\
\theta^{(3)}_3&=&\frac{1}{6}\left( -\alpha_3
(a_1^2+a_2^2+b_1^2+b_2^2)+2\alpha_1a_2b_2 + 2\alpha_2a_1b_1
\right) .\nonumber \\
&& \eqn{theta3}
\end{eqnarray}

\end{itemize}

The special cases $\alpha_1=\alpha_2=\alpha_3$,
$\alpha_1=\alpha_2>\alpha_3$ and $\alpha_1>\alpha_2=\alpha_3$
provide different (and rather more complicated) solutions for the
$\lambda^{(3)}_j$ compared with above. Arriving at the solution in
these cases requires solving up to $f^{(10)}=0$. We will not write
out these results explicitly.

Thus we have
\begin{equation} \label{eq:lazy-inter}
\thetavec(t) = \alphavec t + \thetavec^{(3)}t^3 + O(t^4)
\end{equation}
for $t$ in some interval $[0,\epsilon]$. $\thetavec^{(3)}$ is
given in \eq{theta3}, except in the special cases noted above.

\subsection{Conditions for laziness}

{} From section II.C of \cite{Hammerer02a}, the expression for $C_H(U)$ takes
a simpler form when we have $\theta_1+|\theta_3|\le
\frac{\pi}{4}$. In this special case, $C_H(U)$ is given by the
minimum value of $t_s$ such that
\begin{equation}
\thetavec \prec_s \alphavec t_s,    \eqn{mints}
\end{equation}
where again $\thetavec$ is the vector of canonical-form parameters
of $U$ and $\alphavec$ is the vector of canonical-form parameters
of $H$. This special case certainly holds for the canonical-form
parameters of $e^{-iHt}$ when $t$ is sufficiently small. In this
case \eq{mints} is equivalent to
\begin{eqnarray}
\theta_1 &\le& \alpha_1 t_s \nonumber \\
\theta_1 + \theta_2 - \theta_3 &\le& (\alpha_1 +\alpha_2 - \alpha_3) t_s \nonumber \\
\theta_1 + \theta_2 + \theta_3 &\le& (\alpha_1 +\alpha_2 +
\alpha_3) t_s ,
\end{eqnarray}
which is equivalent to
\begin{eqnarray}
\frac{\theta_1}{\alpha_1} &\le&  t_s \nonumber \\
\frac{\theta_1 + \theta_2 - \theta_3}{\alpha_1 +\alpha_2 - \alpha_3} &\le&  t_s \nonumber \\
\frac{\theta_1 + \theta_2 + \theta_3}{\alpha_1 +\alpha_2 +
\alpha_3} &\le&  t_s .
\end{eqnarray}
Thus,
\begin{equation}
C_H(U)=\max \left\{ \frac{\theta_1}{\alpha_1} ,\frac{\theta_1 +
\theta_2 - \theta_3}{\alpha_1 +\alpha_2 - \alpha_3},\frac{\theta_1
+ \theta_2 + \theta_3}{\alpha_1 +\alpha_2 + \alpha_3}
 \right\}.
\end{equation}
Given Eq.~(\ref{eq:lazy-inter}), we have
\begin{equation}
\tau(t)=t +  \tau^{(3)}t^3 + O(t^4)
\end{equation}
for small $t$, where $\tau^{(3)}$ is given by
\begin{equation}
\tau^{(3)}=\max\left\{ \frac{\theta^{(3)}_1}{\alpha_1}
,\frac{\theta^{(3)}_1 + \theta^{(3)}_2 - \theta^{(3)}_3}{\alpha_1
+\alpha_2 - \alpha_3},\frac{\theta^{(3)}_1 + \theta^{(3)}_2 +
\theta^{(3)}_3}{\alpha_1 +\alpha_2 + \alpha_3}  \right\}.
\end{equation}
It is clear that whenever $\tau^{(3)}<0$, the Hamiltonian is lazy.  It
is also clear that $\tau^{(3)}$ is never greater than zero, because
that would imply $\tau(t)>t$, a contradiction. We find below the
solutions (in terms of the parameters of the Hamiltonian) for
$\tau^{(3)}=0$; all Hamiltonians which do not belong to this solution
set are guaranteed to be lazy. Note, however, that the complement of
this solution set does not entirely characterise the class of lazy
Hamiltonians, since there may be Hamiltonians in this set that are
lazy due to higher-order Taylor coefficients that are negative.  So
our results may not fully characterise the set of {\em all} lazy
Hamiltonians.

Let
\begin{eqnarray}
B_1&=&\theta^{(3)}_1, \\
B_2&=&\theta^{(3)}_1 +
\theta^{(3)}_2 - \theta^{(3)}_3, \\
B_3&=&\theta^{(3)}_1 + \theta^{(3)}_2 + \theta^{(3)}_3.
\end{eqnarray}
The coefficient $\tau^{(3)}$ is zero if and only if at least one of
$B_j$ are zero. It is straightforward to show that for
$\alpha_1>\alpha_2>\alpha_3$,
\begin{eqnarray}
B_1=0  &\Leftrightarrowsp&  a_2=a_3=b_2=b_3=0 \\
B_2 =0&\Leftrightarrowsp&a_1=-b_1; a_2=-b_2; a_3=b_3\\
B_3 =0&\Leftrightarrowsp& a_1=b_1; a_2=b_2; a_3=b_3.
\end{eqnarray}
We have arrived at the main result of this paper:
\begin{result}
Any Hamiltonian of the form of \eq{hs} for which
$\alpha_1>\alpha_2>\alpha_3$ and for which none of the three
conditions
\begin{enumerate}
\item \hspace{0.5cm} $a_2=a_3=b_2=b_3=0$
\item \hspace{0.5cm} $a_1=-b_1; a_2=-b_2; a_3=b_3$
\item \hspace{0.5cm} $a_1=b_1; a_2=b_2; a_3=b_3$
\end{enumerate}
hold, is lazy. Such Hamiltonians will therefore need to be applied
infinitely many times when used in a time-optimal simulation of a
nonlocal two-qubit unitary.
\end{result}

These conditions obviously make it very easy to generate examples of
lazy Hamiltonians, and imply that almost all two-qubit Hamiltonians
are lazy.  Note that the special cases $\alpha_1=\alpha_2>\alpha_3$,
$\alpha_1>\alpha_2=\alpha_3$ and $\alpha_1=\alpha_2=\alpha_3$ yield
somewhat more complicated conditions for a Hamiltonian to be lazy.
These conditions are complex and not very illuminating, but can be
obtained using techniques similar to those described above, so we will
not reproduce them here.

\section{Using lazy Hamiltonian in finite time steps}
\label{sec:finite}

The results of the previous section show that almost all two-qubit
Hamiltonians are lazy. This means that, in a simulation circuit,
infinitesimal time steps must be employed to achieve
time-optimality. We now show that, despite this requirement, if
finite time steps are used then the corresponding sacrifice of
interaction time is not very large --- only a small relaxation
from strict time-optimality is required in order to reduce the
number of time steps to something practical.

To make our results concrete, we consider the case where the unitary
being simulated is the controlled-not ({\sc cnot}) gate. Similar
conclusions can be reached in the general case by following a similar
argument to that below, and making use of the results
of~\cite{Zhang02a}.  It can be shown \cite{Hammerer02a} that the
minimum time for simulating a {\sc cnot} is $C_H(\text{{\sc
    cnot}})=\frac{\pi}{4\alpha_1}$ where $\alpha_1$ is the largest
canonical-form parameter of the interaction Hamiltonian.  When $H$ is
lazy, can we construct a simulation using a finite number of time
steps such that the total interaction time is not much larger than the
optimum $C_H(\text{{\sc cnot}})$? Such a scheme is given in
\cite{Bremner02a}, whereby an arbitrary nonlocal two-qubit unitary $U$
is applied a finite number of times together with local unitaries to
simulate a {\sc cnot}.  Using the scheme in~\cite{Bremner02a}, if $U$
has largest canonical-form parameter $\theta_1$ such that
\begin{equation}
n=\frac{\pi}{4\theta_1}
\end{equation}
is an integer greater than one, then the scheme can be used to
simulate a {\sc cnot} by applying $U$ exactly $n$ times. Of
course, we are interested in the case when $U=e^{-iH\Delta}$, that
is $U$ is given by the evolution of an interaction Hamiltonian
over a time $\Delta$. The total interaction time would then be
\begin{equation}
t_s=n\Delta=\frac{\pi\Delta}{4\theta_1(\Delta)}. \eqn{ts}
\end{equation}
{} From the previous section, the function $\theta_1(\Delta)$ can be
written, for small $\Delta$, as
\begin{equation}
\theta_1(\Delta)=\alpha_1\Delta +
\theta^{(3)}\Delta^3+O(\Delta^4).
\end{equation}
Thus, for small $\Delta$,
\begin{eqnarray}
t_s&=&\frac{\pi}{4\alpha_1 + \theta_1^{(3)}\Delta^2 + O(\Delta^3)}
\\
 &=& C_H(\text{{\sc cnot}}) - \frac{\pi \theta_1^{(3)} \Delta^2}{16 \alpha_1^2}
+ O(\Delta^3).
\end{eqnarray}
This shows that to simulate a {\sc cnot} gate by applying a lazy
interaction Hamiltonian in a (finite) number of small time-steps, then
the penalty in the total interaction time, as compared with the
optimum, is only of order $\Delta^2$.

\begin{figure}
\begin{center}
\epsfxsize=9cm \epsfbox{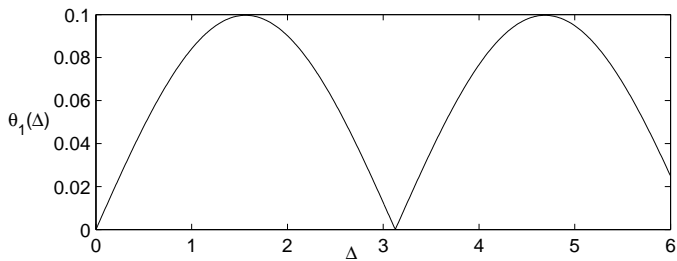} \caption{ Canonical-form
parameter $\theta_1(\Delta)$ of the unitary $e^{-iH\Delta}$, where
$H = 0.1 X \otimes X+ I \otimes Z$.
}\label{fig:numeric}
\end{center}
\end{figure}

\begin{figure}
\begin{center}
\epsfxsize=9cm \epsfbox{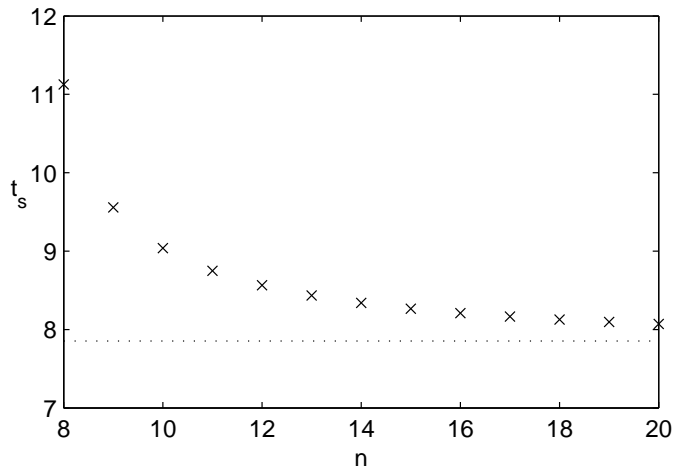} \caption{ Total interaction
time as a function of the number of simulation steps, for the
simulation of {\sc CNOT} using the Hamiltonian $H=0.1X\otimes X +
I\otimes Z$. }\label{fig:example}
\end{center}
\end{figure}

As an example, consider a specific interaction Hamiltonian $H=0.1
X\otimes X + I\otimes Z$. Using the results of the previous section it
can easily be verified that $H$ is lazy.  The graph of
$\theta_1(\Delta)$ as a function of $\Delta$ is shown in
Fig.~\ref{fig:numeric}.  We choose a range of positive integer values
of $n$, and for each $n$ we calculate how long the corresponding time
step ($\Delta$) is by numerically solving
\begin{equation}
\theta_1(\Delta)=\frac{\pi}{4n}. \eqn{solveit}
\end{equation}
No solution to \eq{solveit} exists for $n<8$.  This can be seen from
the fact that $\pi/(4\times 7)=0.112\dots$, which is greater than the
maximum value that $\theta_1(\Delta)$ takes.  For $n$ equal to 8 or
greater, a corresponding $\Delta$ can be found.  Finally, the
interaction time $t_s$ required to simulate the {\sc cnot} is
calculated via \eq{ts}. The results are shown in
Fig.~\ref{fig:example}. The dashed line is the optimal time,
$C_H(\text{{\sc cnot}})=5\pi/2$. The results clearly show a
near-optimal simulation with relatively small numbers of time steps.
For 20 time steps, the total interaction time is just $2.8$ percent
greater than the optimal.

\section{Conclusions} \label{se:future}

We have defined a class of ``lazy'' two-qubit Hamiltonians, those
which can simulate themselves faster with the aid of fast local
control than with uninterrupted evolution. When a lazy Hamiltonian
is used in the time-optimal simulation of any nonlocal two-qubit
unitary, we have shown that the simulation will require an
infinite number of steps, and thus will be impractical. We have
derived a simple set of sufficient conditions enabling us to prove
that a given Hamiltonian is lazy.  This set of conditions implies
that almost all two-qubit Hamiltonians are lazy.  Finally, we have
shown that only a rather small sacrifice in the simulation time
needs to be made in order to use a lazy Hamiltonian in a
finite-step simulation.

\acknowledgments

Thanks to Chris Dawson for helpful discussions and for suggesting
the name ``lazy'' Hamiltonian.  Thanks also to Guifr\'e Vidal for
enlightening discussions.  HLH and MAN enjoyed the hospitality of
the Institute for Quantum Information at the California Institute
of Technology, where part of this work was completed.


\end{document}